\documentclass[epj,twocolumn]{webofc}
\usepackage[varg]{txfonts}   % Web of Conferences font
\usepackage{graphicx}

\wocname{epj}
\woctitle{Seismology of the Sun and the Distant Stars 2016}

\newcommand{\apjs}{APJS\ }

\newcommand{\mnras}{MNRAS\ }
\newcommand{\aap}{A\&A\ }
\newcommand{\apjl}{APJL\ }
\newcommand{\msun}{{\rm M_\odot}}
\newcommand{\almlt}{\alpha_{\rm MLT}}
\newcommand{\xini}{X_{\rm ini}}
\newcommand{\yini}{Y_{\rm ini}}
\newcommand{\zini}{Z_{\rm ini}}
\newcommand{\dnu}{\Delta \nu}
\newcommand{\numax}{\nu_{\rm max}}
\newcommand{\teff}{T_{\rm eff}}
\newcommand{\logg}{\log{g}}
\newcommand{\feh}{{\rm [Fe/H]}}
\newcommand{\bespp}{\texttt{BeSPP}}

\begin{document}
%***********************************************************************
\title{The more the merrier: grid based modelling of Kepler dwarfs with 5-dimensional stellar grids}

\author{\firstname{Aldo} \lastname{Serenelli}\inst{1}\fnsep\thanks{\email{aldos@ice.csic.es}} \and
        \firstname{William J.} \lastname{Chaplin}\inst{2,3} \fnsep \and
        \firstname{Daniel} \lastname{Huber\inst{3,4,5,6} }}

\institute{Institute of Space Sciences (IEEC-CSIC), Campus UAB, Barcelona, E-08193, Spain
\and
           School of Physics and Astronomy, University of Birmingham, Edgbaston, Birmingham B15 2TT, United Kingdom 
           \and 
           	Stellar Astrophysics Centre, Dept. of Physics and Astronomy, Aarhus Univ., Ny Munkegade 120, DK-8000 Aarhus C, Denmark
           \and 
           Institute for Astronomy, University of Hawaii, 2680 Wood- lawn Drive, Honolulu, HI 96822, US
           \and Sydney Institute for Astronomy, School of Physics, University of Sydney, Sydney, Australia
           \and
           SETI Institute, 189 Bernardo Avenue, Mountain View, CA 94043  }

%-----------------------------------------------------------------------
\abstract{%
We present preliminary results of our grid based modelling (GBM) of the dwarf/subgiant sample of stars observed with Kepler including global asteroseismic parameters. GBM analysis in this work is based on a large grid of stellar models that is characterized by five independent parameters: model mass and age, initial metallicity ($\zini$), initial helium ($\yini$), and mixing length parameter ($\alpha_{mlt}$). Using this grid relaxes assumptions used in all previous GBM work where the initial composition is determined by a single parameter and that $\alpha_{mlt}$ is fixed to a solar-calibrated value. The new grid allows us to study, for example, the impact of different galactic chemical enrichment models on the determination of stellar parameters such as mass radius and age. Also, it allows to include new results from stellar atmosphere models on $\alpha_{mlt}$ in the GBM analysis in a simple manner. Alternatively, it can be tested if global asteroseismology is a useful tool to constraint our ignorance on quantities such as $\yini$ and $\alpha_{mlt}$. Initial findings show that mass determination is robust with respect to freedom in the latter quantities, with a 4.4\% maximum deviation for extreme assumptions regarding prior information on $\yini-\zini$ relations and $\alpha_{mlt}$. On the other hand,  tests carried out so far seem to indicate that global seismology does not have  much power to constrain $\yini-\zini$ relations of $\alpha_{mlt}$ values without resourcing to additional information.
}
\maketitle
%
%-----------------------------------------------------------------------
\section{Introduction}
\label{intro}
Grid based modelling (GBM) is now routinely employed in the determination of stellar parameters using global asteroseismic quantities, i.e. the large frequency separation $\dnu$ and the frequency of maximum power $\numax$ (see e.g. \cite{chaplin:2014}). Many quantities need to be determined to construct the large grids of stellar models needed for GBM. Leaving aside variations in the input physics employed in stellar evolution calculations, the most important simplifications made in order to keep the problem tractable are the assumption of a fixed mixing length parameter $\almlt$ and a uniquely relation between the initial helium $\yini$ and metallicity $\zini$ of the models. The latter, together with the normalization $\xini + \yini + \zini =1$ and an adopted element mixture leave only one free parameter associated with the initial composition of models. This sets the minimal framework for GBM in which each stellar model in the grid is determined by three independent parameters: its age, the initial mass of the evolutionary sequence to which the stellar model belongs (might differ from its actual mass due to mass loss) and the initial $\zini$ (or any other parameter defining the initial composition). Needless to say, such construction of models inherently include strong prejudices: that a solar $\almlt$ is universal and that $\yini$ and $\zini$ are uniquely and universally related to each other. These are strong and limiting assumptions.

In the framework of GBM the way to break free is to use stellar grids where $\almlt$ is also a free parameter, together with a second parameter linked to initial composition. Based on this necessity, in this work we present GBM results based on a newly computed grid of stellar models based on a five-dimensional parameter space: age, mass (mass loss is negligible for dwarfs and subgiant stars), $\zini$, $\yini$, and $\almlt$ are the parameters that characterize a given stellar model in the grid. The grid of models is combined with \bespp, a Bayesian algorithm \cite{serenelli:2013} developed to determine stellar physical quantities from any appropriate combination of  spectroscopic, photometric or asteroseismic quantities. Different assumptions regarding galactic chemical enrichment, e.g. by introducing a relation between $\yini$ and $\zini$, or regarding dependences of $\almlt$ on surface stellar parameters ($\teff$, $\logg$ and $\feh$) can be tested by considering appropriate prior probabilities in \texttt{BeSPP}. We apply our methods to the Kepler sample from \cite{chaplin:2014} in order to answer the following questions. How  robust is the determination of stellar parameters given our lack of knowledge regarding $\yini$ and $\almlt$ in particular? Inverting the problem,  can global seismology be used to establish a history, e.g. in the form of a $\yini-\zini$ relation, of galactic chemical enrichement? Can it be used to empirically establish a relation between $\almlt$ and stellar properties to, e.g. test recent results from three dimensional stellar atmosphere models \cite{magic:2015}?

%-----------------------------------------------------------------------
\section{Methods and results}
\label{sec-1}

\begin{figure}[h]
\centering
\includegraphics[scale=.1,clip]{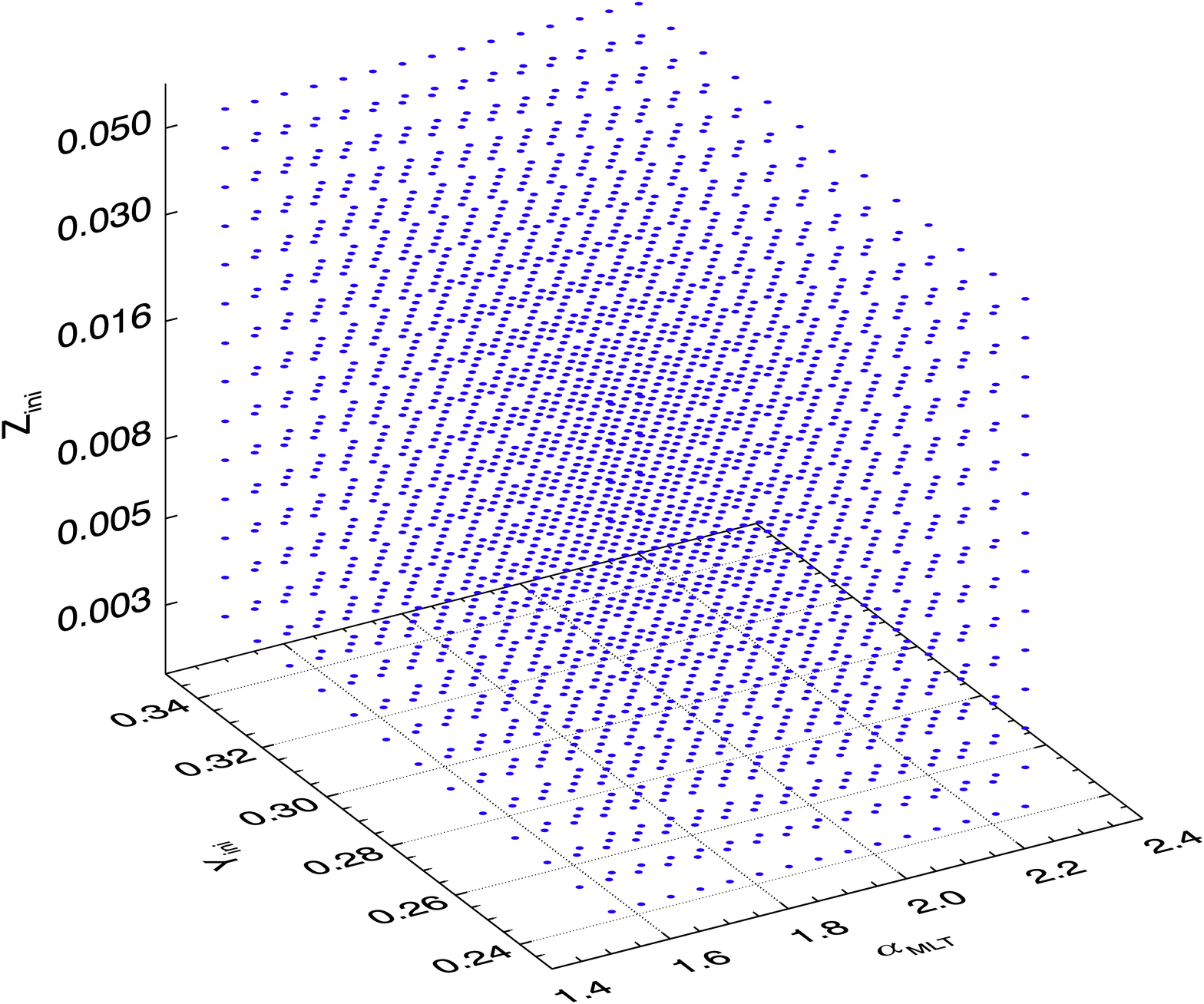}
\caption{Initial composition and $\almlt$ parameter space of the grid of stellar models used in the present work.}
\label{fig:grid}       % Give a unique label
\end{figure}

Stellar models have been computed using \texttt{GARSTEC}. For the present work, the grid spans a mass range between 0.7 and 1.8~$\msun$ with $\Delta M/\msun=0.02$ and extends down to $\logg=3.2$. $\yini$ and $\zini$ and $\almlt$ are taken as three independent parameters and the qubic parameter space covered is illustrated in Fig.\,\ref{fig:grid}. Note we take $\zini$, not $\feh$ as independent parameter. However, the grid is uniformly spaced in $\log{\zini}$. 

Depending on the problem at hand, priors are included in \bespp\ that account for different assumptions regarding model parameters. In this work we consider different possibilities regarding priors for initial composition or $\almlt$ as follows: 
\begin{itemize}

\item[-] \textbf{Flat}: flat priors are used for composition and $\almlt$.

\item[-] $\mathbf{\Delta_{{\rm YZ}} = 1.2}$: a Gaussian prior relates $\yini$ and $\zini$ such that
\begin{equation}
p(\yini) \propto \exp{\left[- \frac{(\yini - Y_\Delta)^2}{2 \sigma_Y^2} \right]}, \label{eq:delta}
\end{equation}
with $Y_\Delta = Y_{\rm SBBN} + \Delta \zini$, $\Delta =1.2$ and $\sigma_Y=0.01$. 

\item[-] \textbf{3DMLT}: a Gaussian prior relating $\almlt$ with 3D model atmosphere results such that
\begin{equation}
p(\alpha_{\rm MLT}) \propto \exp{\left[-\frac{(\almlt - \alpha_{\rm 3D})^2}{2\sigma_\alpha^2} \right]}, \label{eq:3d}
\end{equation}
with $\alpha_{\rm 3D} \equiv \alpha_{\rm 3D}(\teff,\logg,\feh)$ from \cite{magic:2015} (but shifted by a constant value to match our solar calibrated value $\alpha_{\rm MLT}= 1.802$) and $\sigma_\alpha$=0.05. 

\item[-] \textbf{3DMLT + $\mathbf{\Delta_{\rm YZ}=1.2}$}: combination of Eqs.\,\ref{eq:delta}~and~\ref{eq:3d}
\begin{equation}
p(\yini,\almlt) \propto p(\yini) \cdot p(\almlt). \label{eq:delta3d}
\end{equation}

\item[-] \textbf{SUN}: effectively constraints parameter space to the standard three dimensional space by using
\begin{equation}
p(\yini) \propto \exp{\left[- \frac{(\yini - Y_\Delta)^2}{2 \sigma_Y^2} \right]} \label{eq:sun}
\end{equation}
as before and the solar calibrated $\almlt=1.802$.
\end{itemize}

Input data in our analysis is taken from \cite{chaplin:2014} and includes $\teff$, $\feh$, $\dnu$ and $\numax$. In that work, $\feh$ values for all stars were taken equal to a mean solar neighborhood value and do not reflect any spectroscopic or photometric determination.

\begin{figure}[h]
\centering
\includegraphics[width=4.3cm,clip]{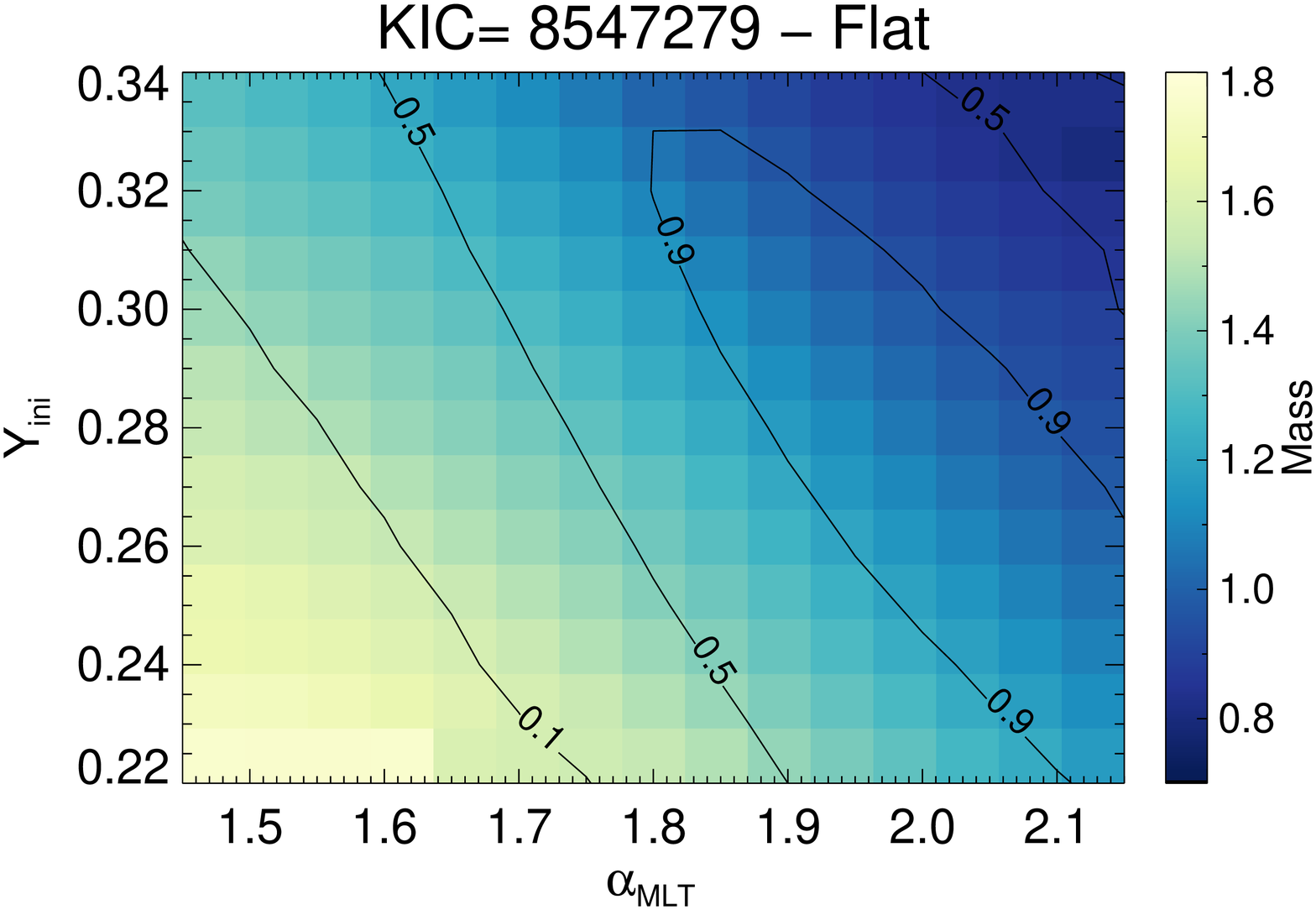}\includegraphics[width=4.3cm,clip]{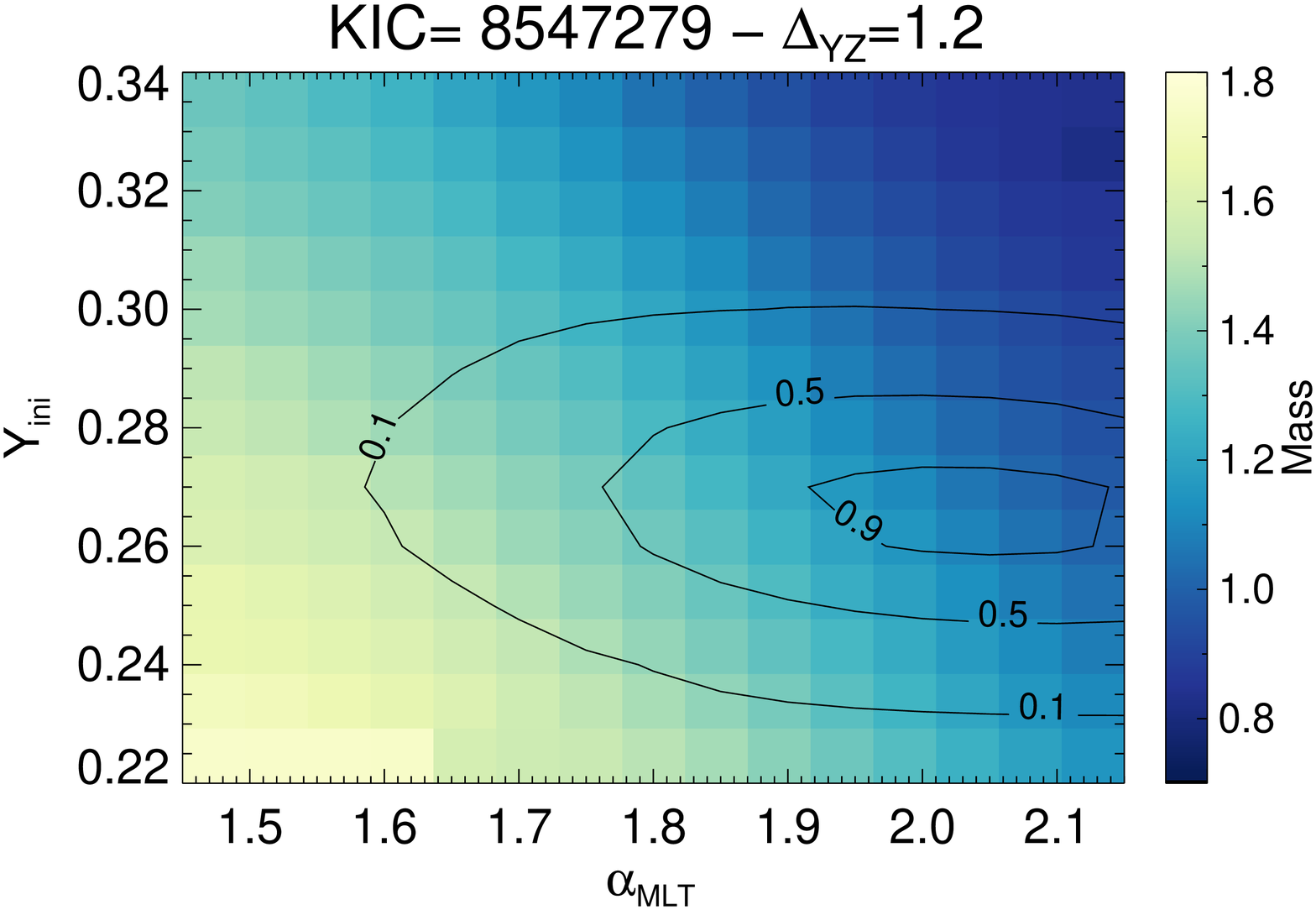}
\includegraphics[width=4.3cm,clip]{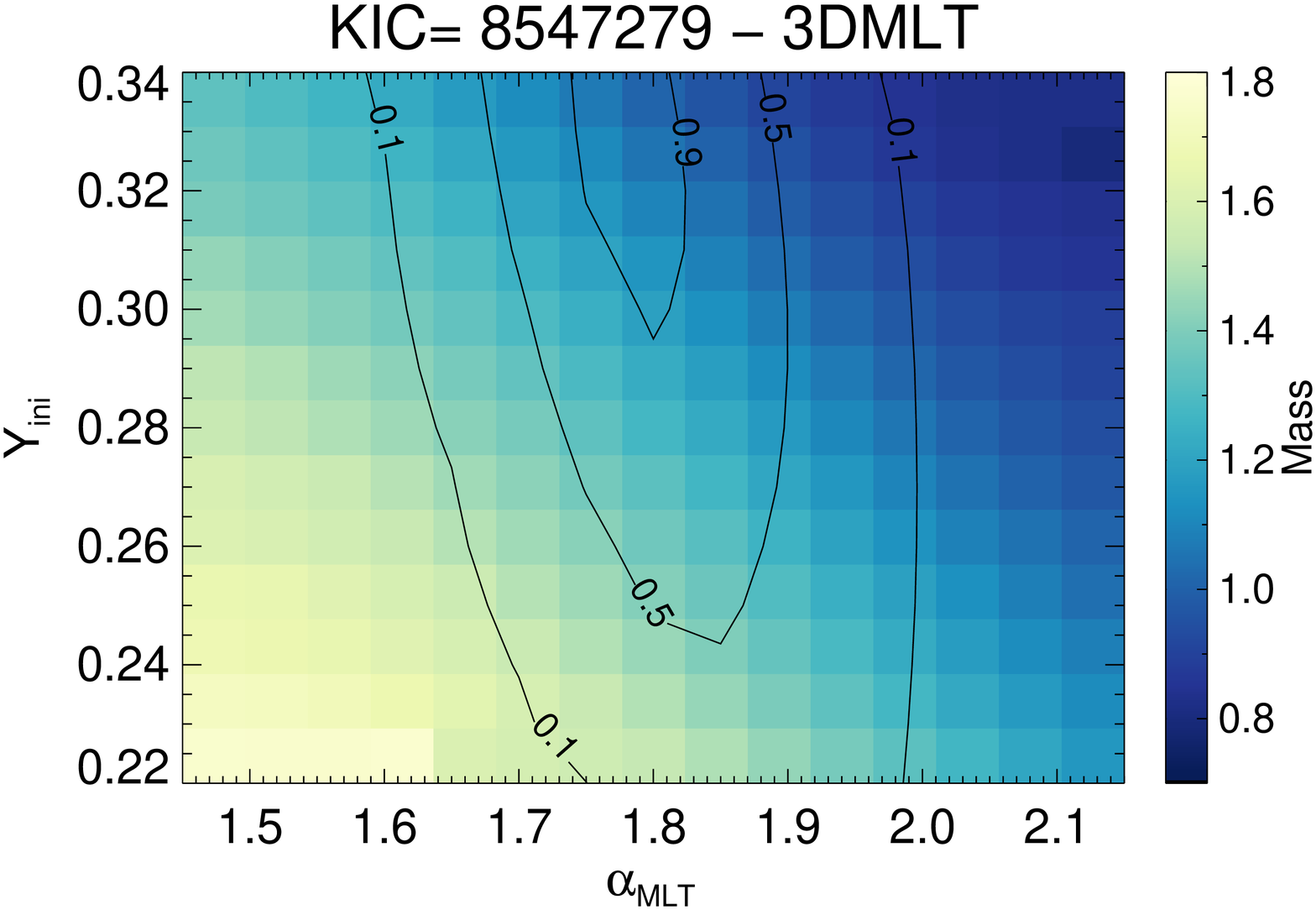}\includegraphics[width=4.3cm,clip]{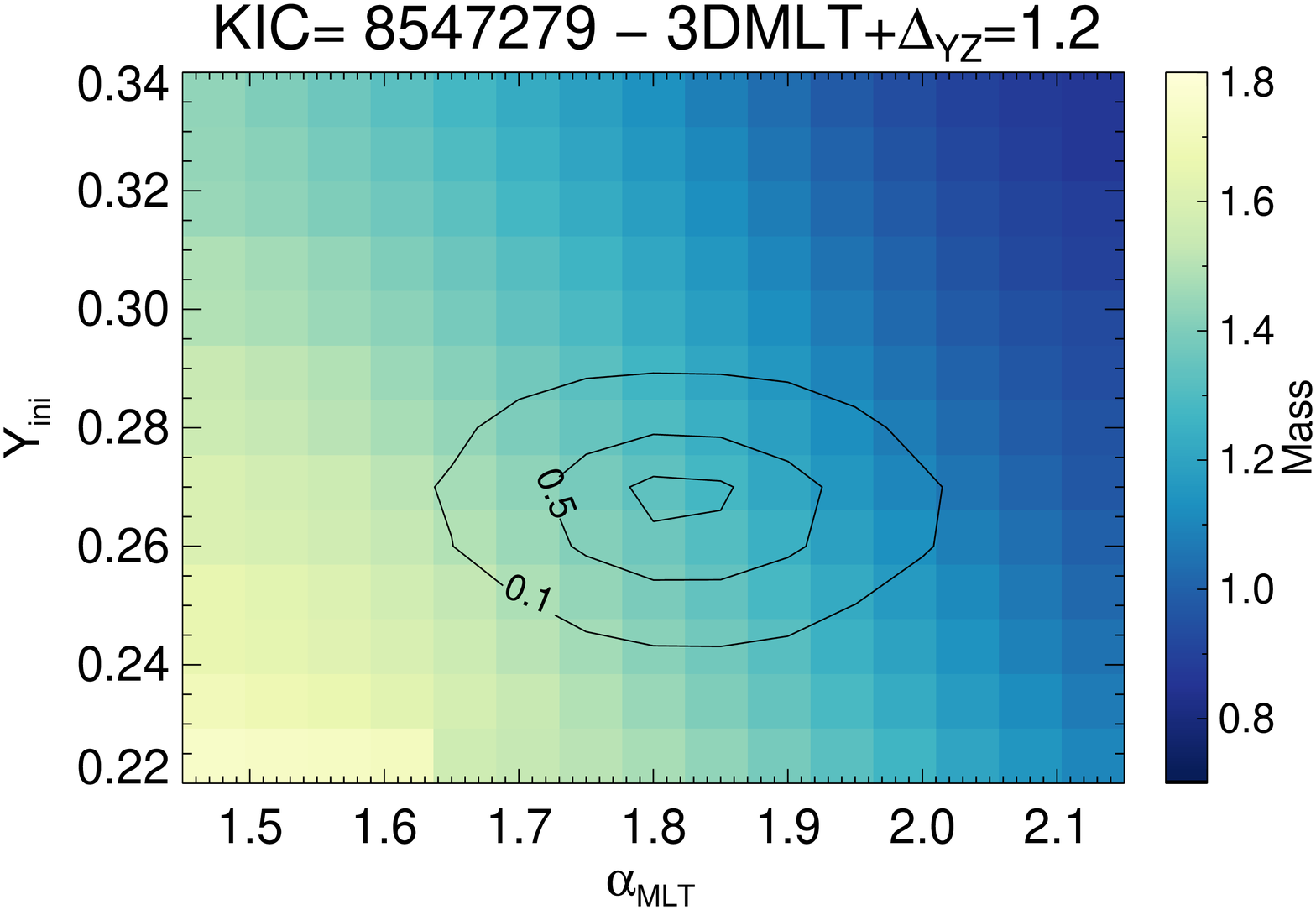}
\caption{Impact of using different priors in the mass determination of KIC 8547279. For each $\yini-\almlt$ point, mass value is the mean value of the mass PDF after marginalizing the 5-dimensional PDF over age and metallicity. Contours depict isoprobability curves for the mass as a function of $\yini$ and $\almlt$.}
\label{fig:kic85}       % Give a unique label
\end{figure}

Fig.\,\ref{fig:kic85} shows results for the mass determination of KIC~8547279 in the $\yini-\almlt$ plane. Mass is color coded, and at each point in the plot, the mass value marginalized over $\zini$, and age is shown. Contours correspond to 0.9, 0.5, and 0.1 probability levels. Priors applied in each case are given in each plot title. Very little constraining power comes from the asteroseismic analysis for $\yini$ and $\almlt$ when flat priors are applied. This is probably due to a degenerate behavior than $\almlt$ and $\yini$ have on $\teff$, as can be seen by the diagonal orientation of the contours. Increasing either $\yini$ or $\almlt$ leads to higher $\teff$ in models, but given that $\teff$ is bound by observations, then our results show a compensation effect: higher $\yini$ correspond to lower $\almlt$ values and viceversa. 

%++++++++++++++
%++++++++++++++

\begin{figure}[h]
\centering
\includegraphics[width=4.3cm,clip]{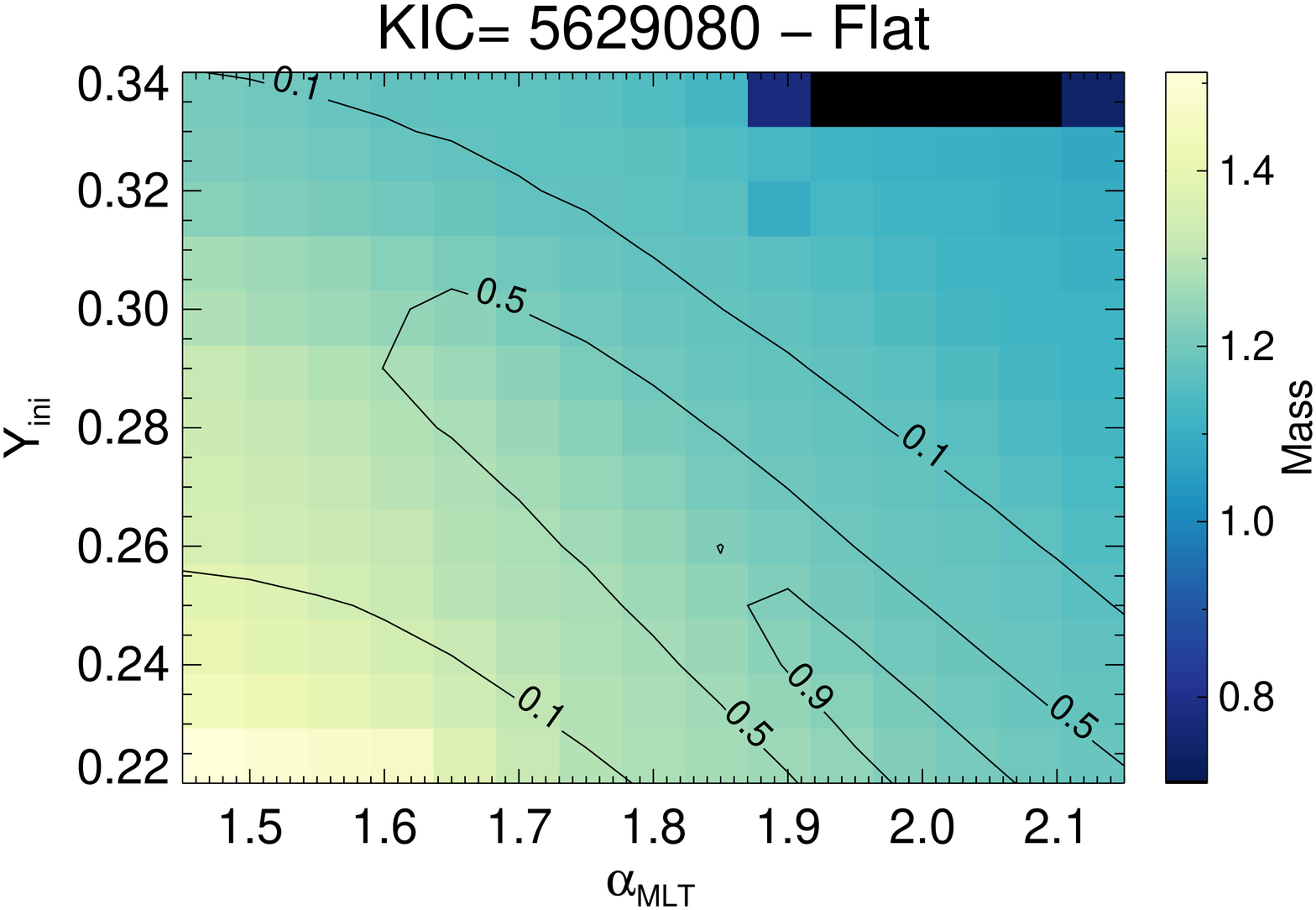}\includegraphics[width=4.3cm,clip]{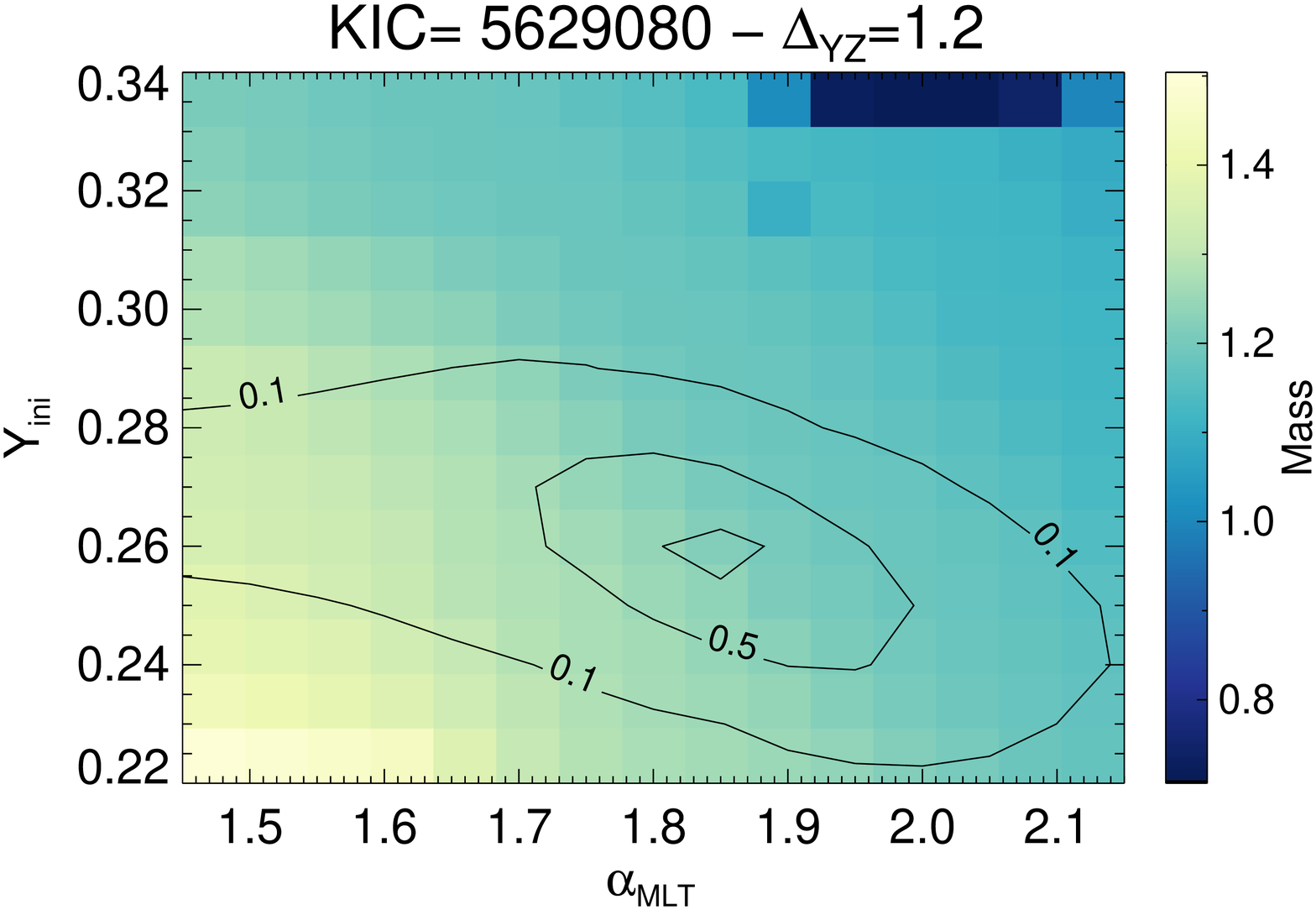}
\includegraphics[width=4.3cm,clip]{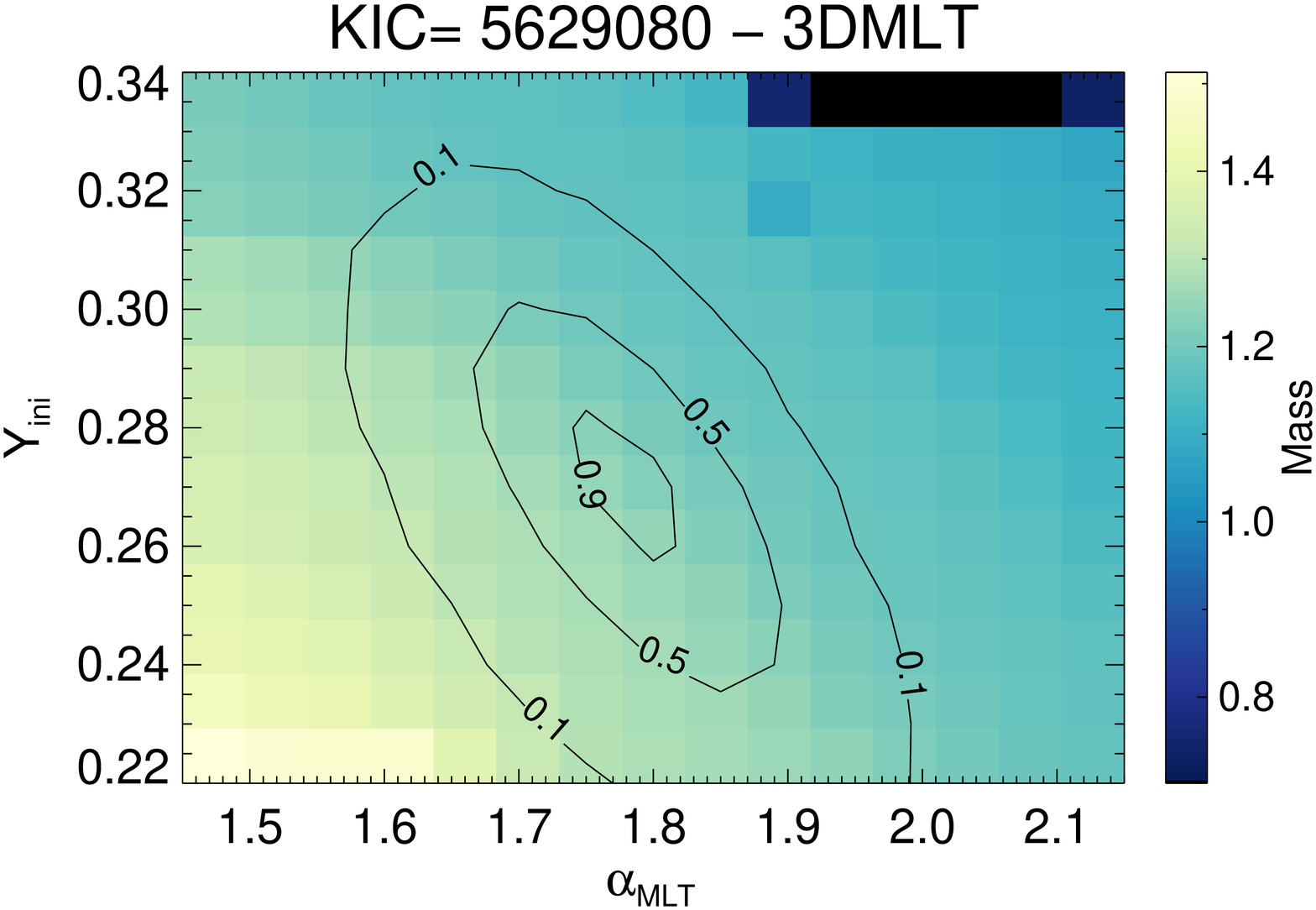}\includegraphics[width=4.3cm,clip]{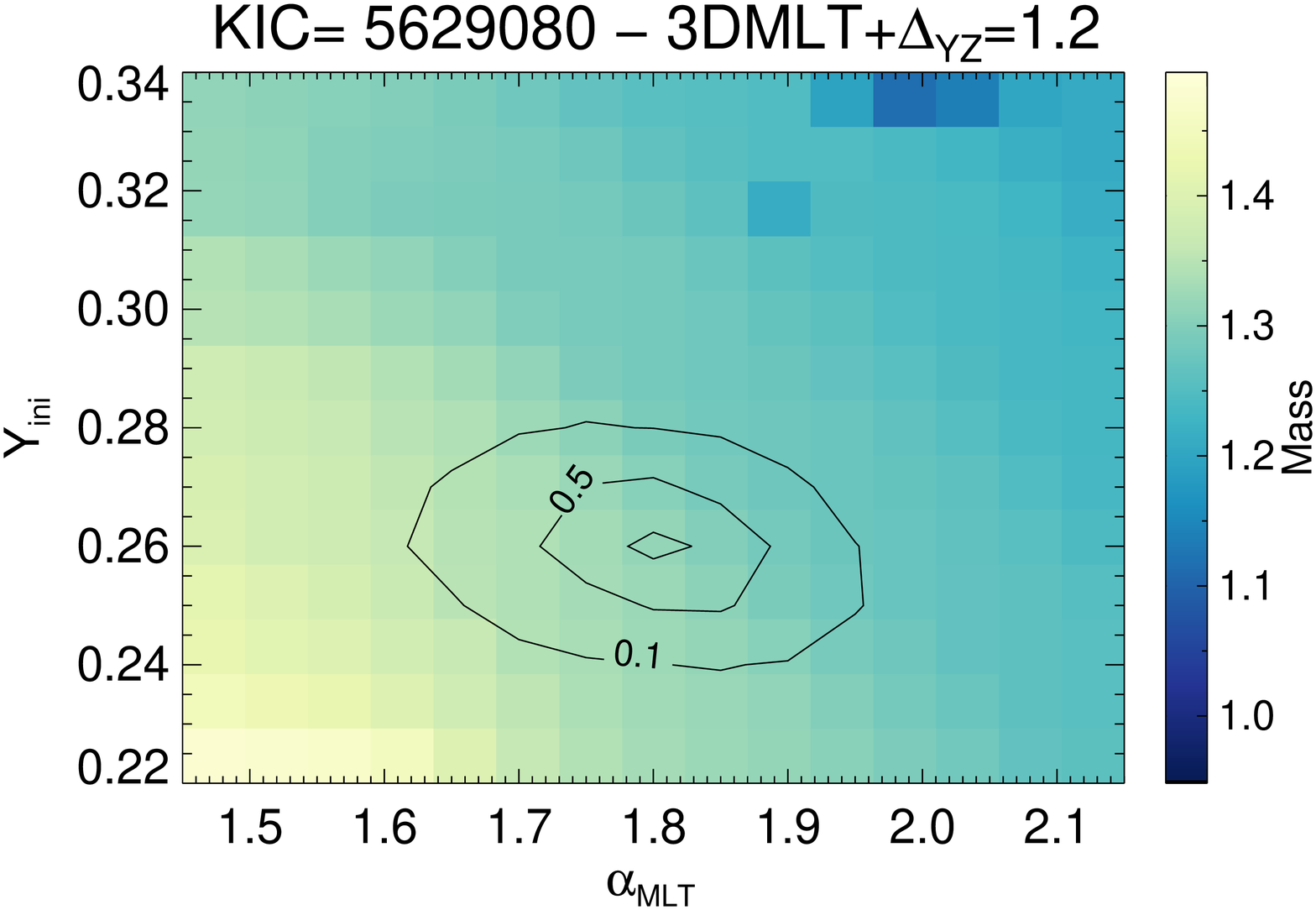}
\caption{Same as Fig.\,\ref{fig:kic85} but for KIC~5629080.}
\label{fig:kic56}       % Give a unique label
\end{figure}

As priors are added, results become more constrained for $\yini$ and $\almlt$, as expected. It is interesting to note, however, that the estimated mass determination is not too sensitive to the choice of priors. In fact, KIC~8547279 is one of the worst cases we have found in this study because the estimated mass varies by about 15\% depending on the priors used. Note that the comparison between the flat and the 3DMLT + $\Delta_{\rm YZ}=1.2$ priors is an extreme one, so this level of variation in mass determinations due to ignorance regarding $\yini$ and $\almlt$ is probably an upper limit. In Fig.\,\ref{fig:kic56} we show one more example, this time for KIC~5629080, which shows almost no variation in its mass estimate. However, as for the previous case, constraints for $\yini$ and $\almlt$ come mostly from priors; asteroseismic data is in fact providing little or no information about these two quantities. 

\begin{figure}[h!]
\centering
\includegraphics[width=4.3cm,clip]{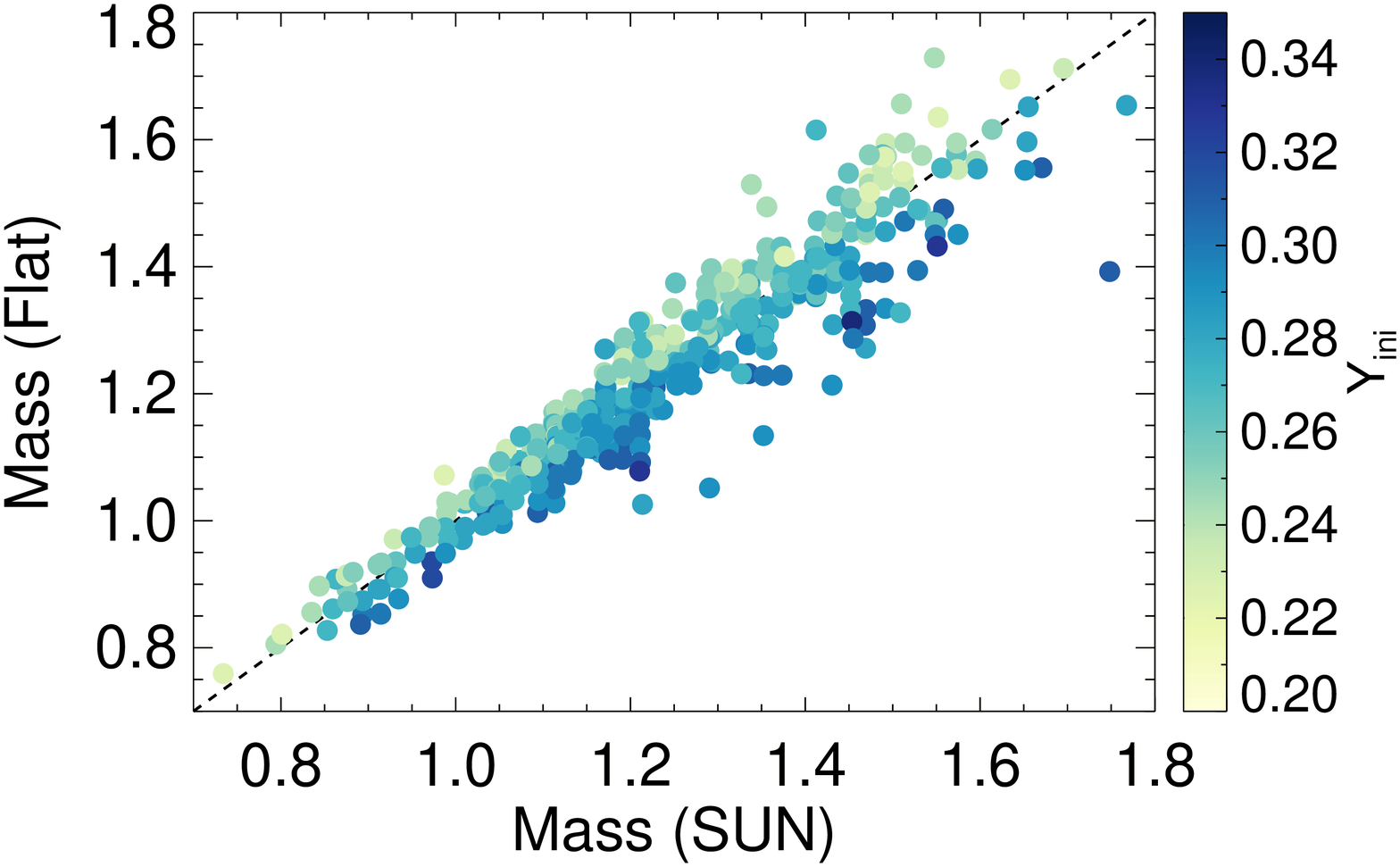}\includegraphics[width=4.3cm,clip]{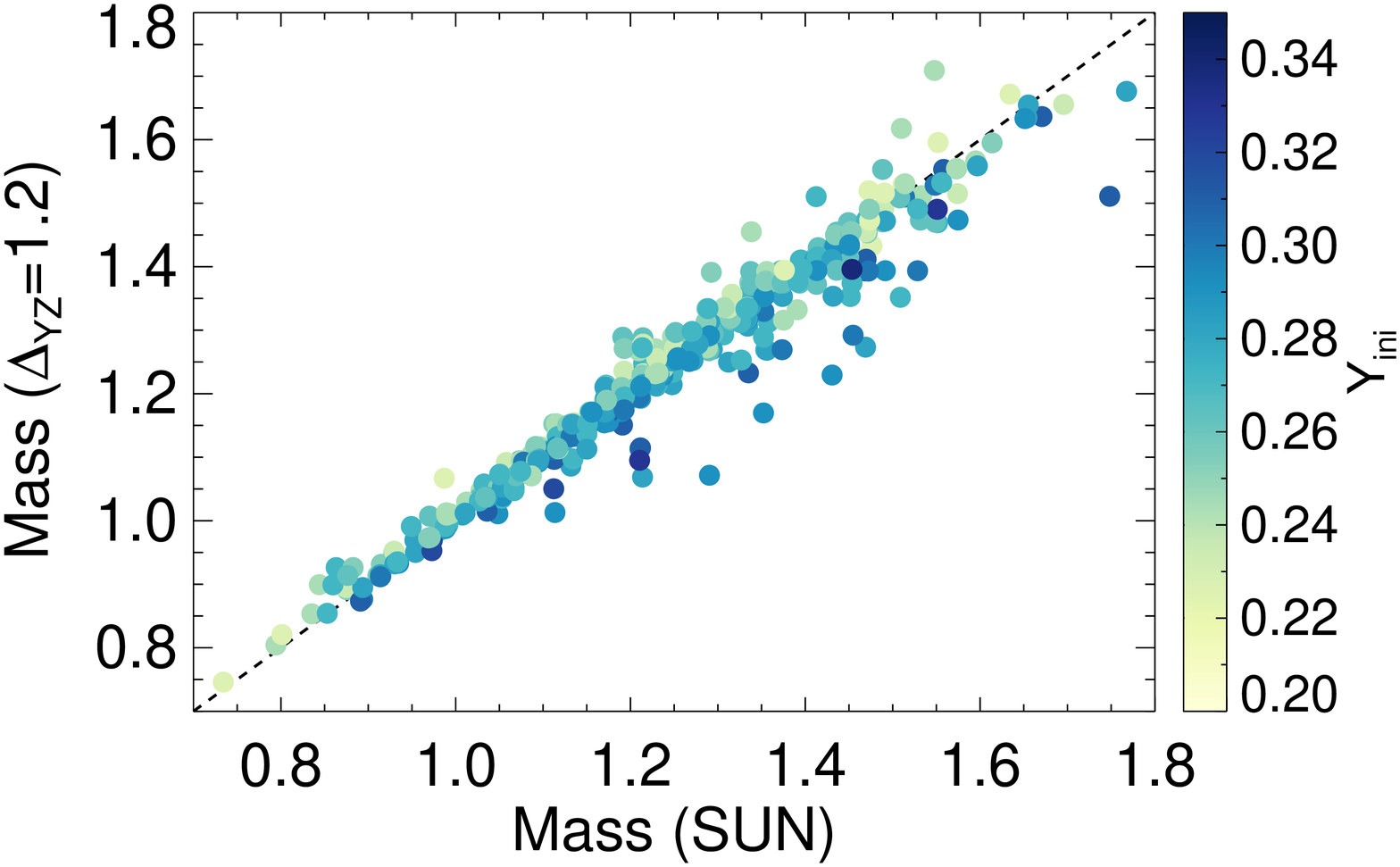}
\includegraphics[width=4.3cm,clip]{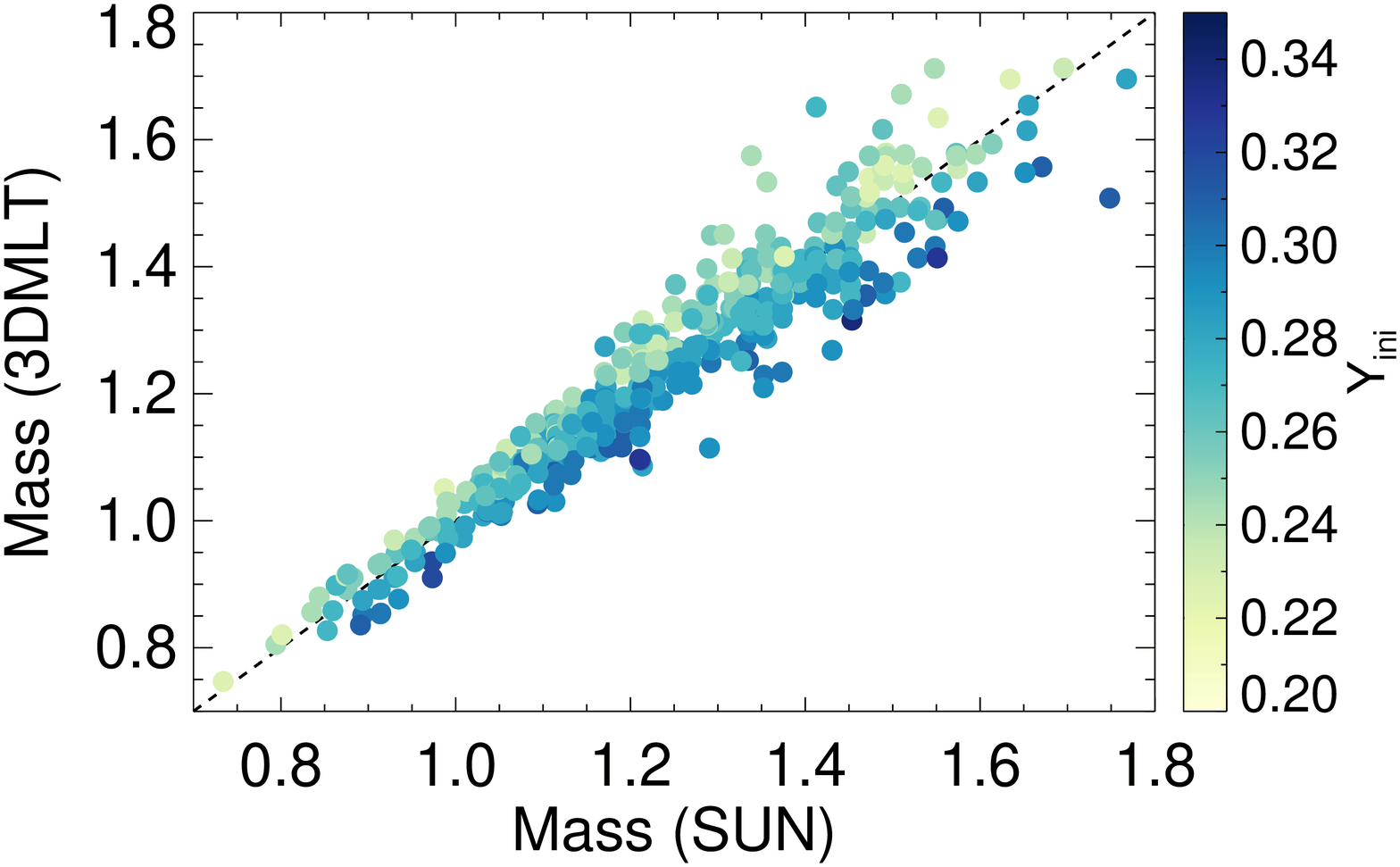}\includegraphics[width=4.3cm,clip]{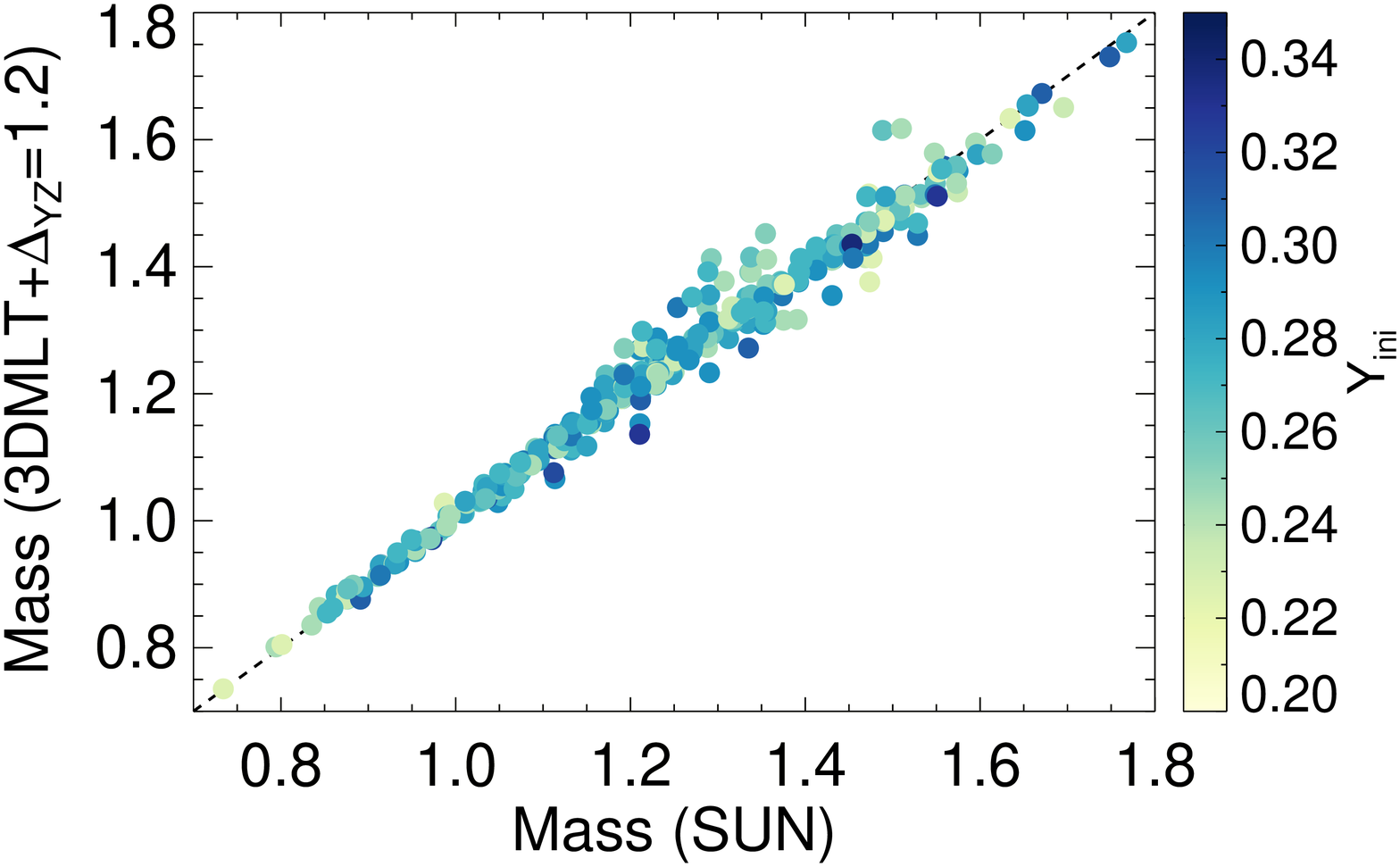}
\caption{Mass determinations for the whole \cite{chaplin:2014} sample depending on the different priors used in \texttt{BeSPP} as indicated in the axes.}
\label{fig:masses}       % Give a unique label
\end{figure}

Fig.\,\ref{fig:masses} summarizes results for the mass determinations in the whole sample by comparing results obtained for the 5 different assumptions on priors given above. In all cases, the x-axis corresponds to priors in Eq.\,\ref{eq:sun}. The maximum dispersion is seen in the comparison Flat-Sun (top left panel) and it is at most 15\% (except for a few cases). In fact, the fractional mass difference between the two cases has $\sigma_{\rm F,\odot}= 4.4\%$, which can be taken as an estimate of the uncertainty between assuming no prior info on initial composition and $\almlt$. We believe this is a very encouraging result. 
The prior on $\almlt$ (bottom left panel) leads to almost the same mass determination as the flat prior, i.e. constraints on $\almlt$ do not affect mass determinations by themselves. On the contrary, relating $\yini$ to $\zini$ (Eq.\,\ref{eq:delta}) influences mass estimates. Interestingly, the interplay between $\yini$ and $\almlt$ due to their degeneracy in relation to $\teff$ values, as described before, make $\almlt$ priors relevant if used in combination with constrained initial composition, as shown by comparision of the upper and lower right panels in Fig.\,\ref{fig:masses}.

%++++++++++++++
%++++++++++++++

Recently, 3-dimensional model atmospheres have been used to determine the dependence of $\almlt$ on stellar surface properties
\cite{trampedach:2014,magic:2015}. But it is not easy to test these results because observational tests have to rely on $\teff$ determinations and, in stellar models, the $\teff$ scale does not depend solely on $\almlt$ but also, for example, on both the metal and helium abundance of stars. Here, we consider whether or not global seismic quantities, in combination with $\teff$ and $\feh$ measurements, can constrain $\almlt$ in the $\teff-\logg$ plane. 

%++++++++++++++
\begin{figure}[h]
\centering
\includegraphics[width=4.3cm,clip]{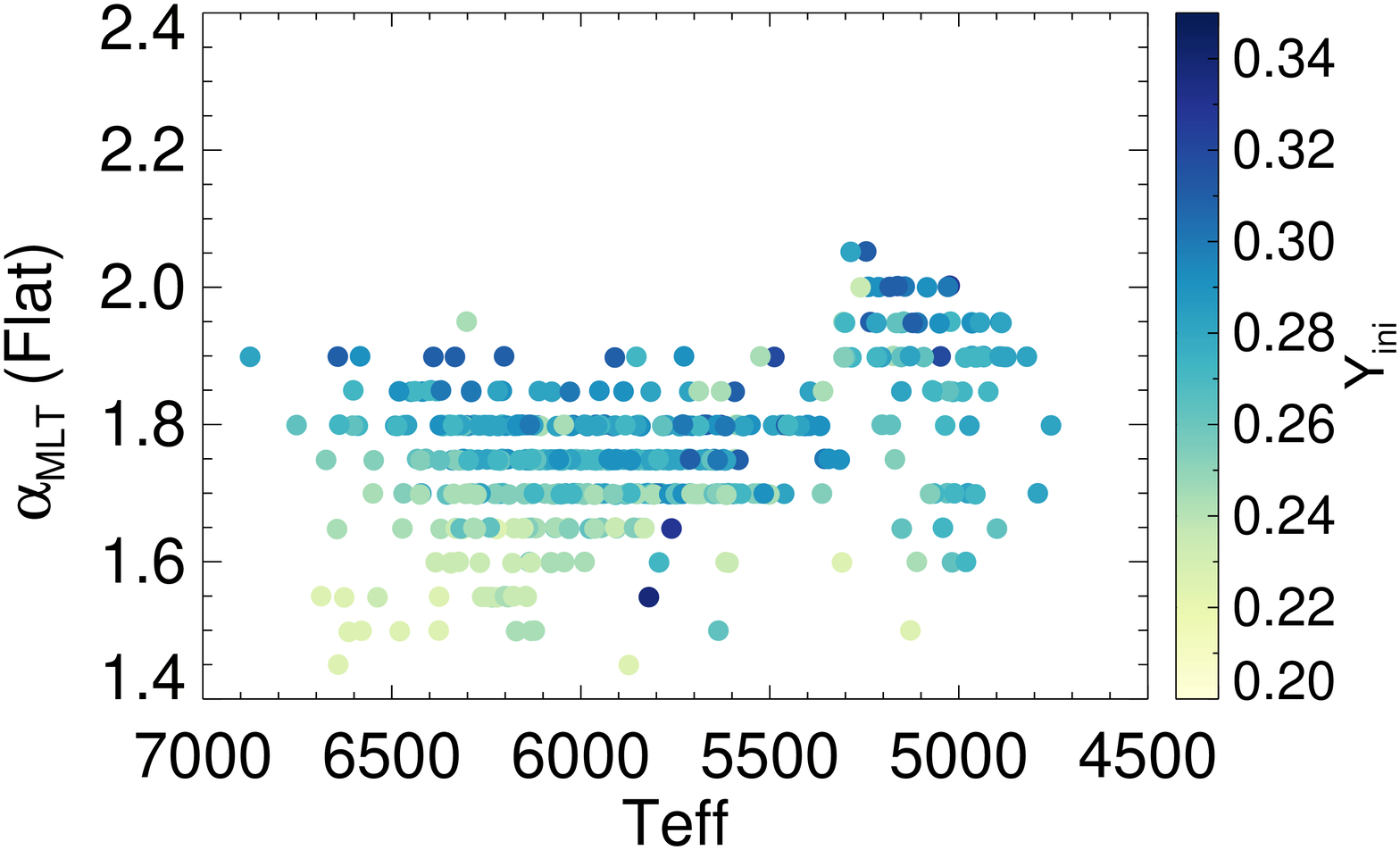}\includegraphics[width=4.3cm,clip]{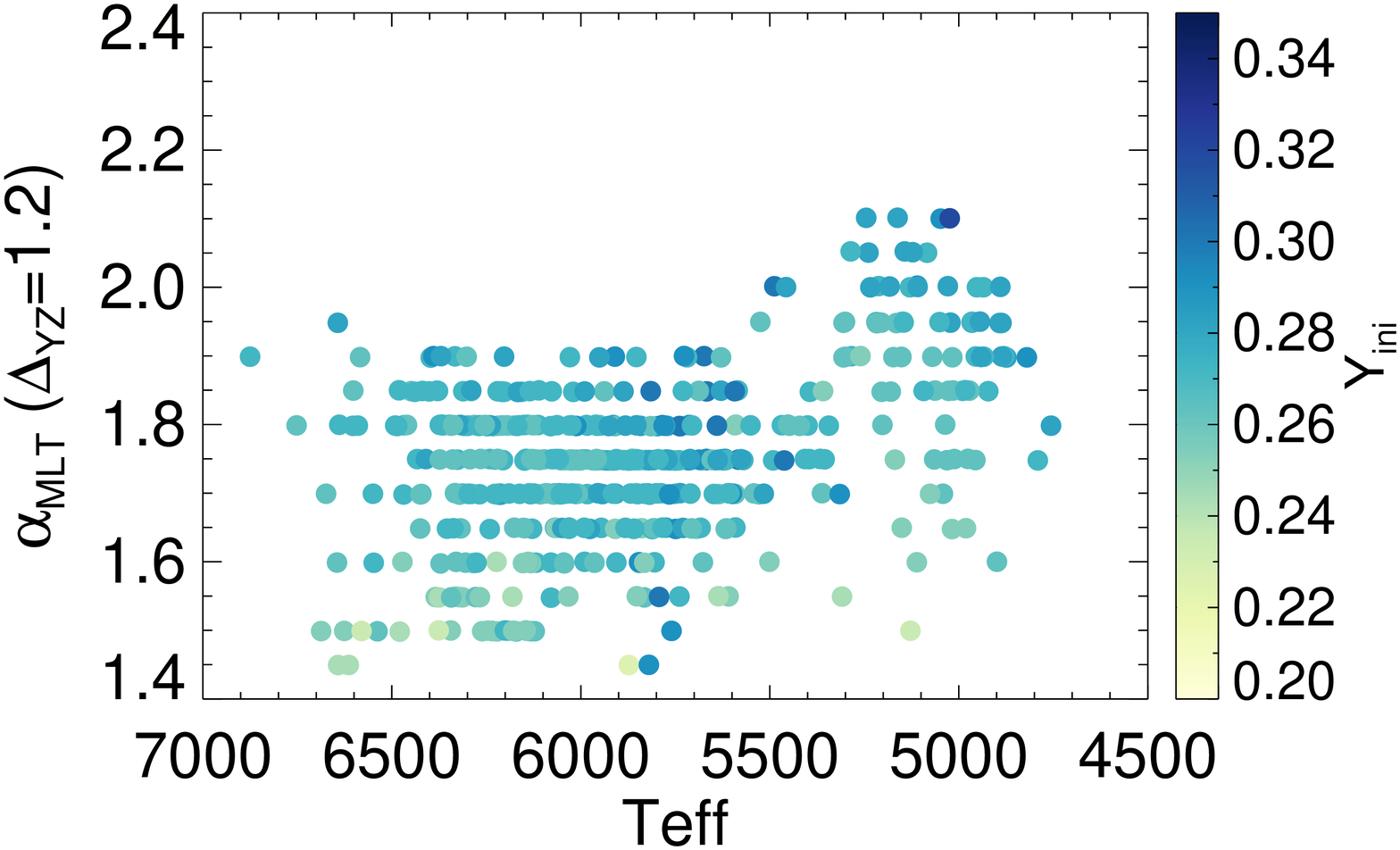}
\includegraphics[width=4.3cm,clip]{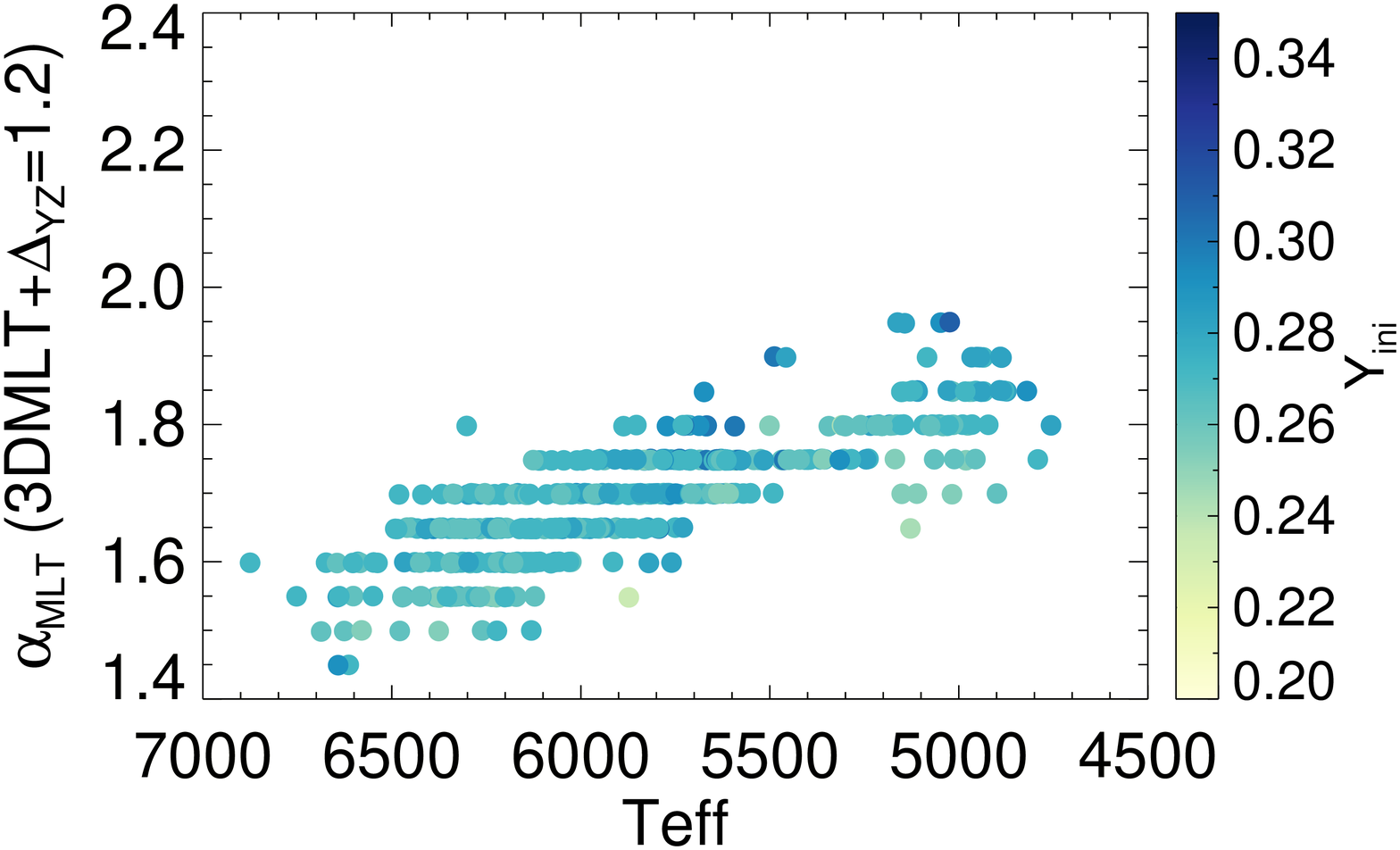}\includegraphics[width=4.3cm,clip]{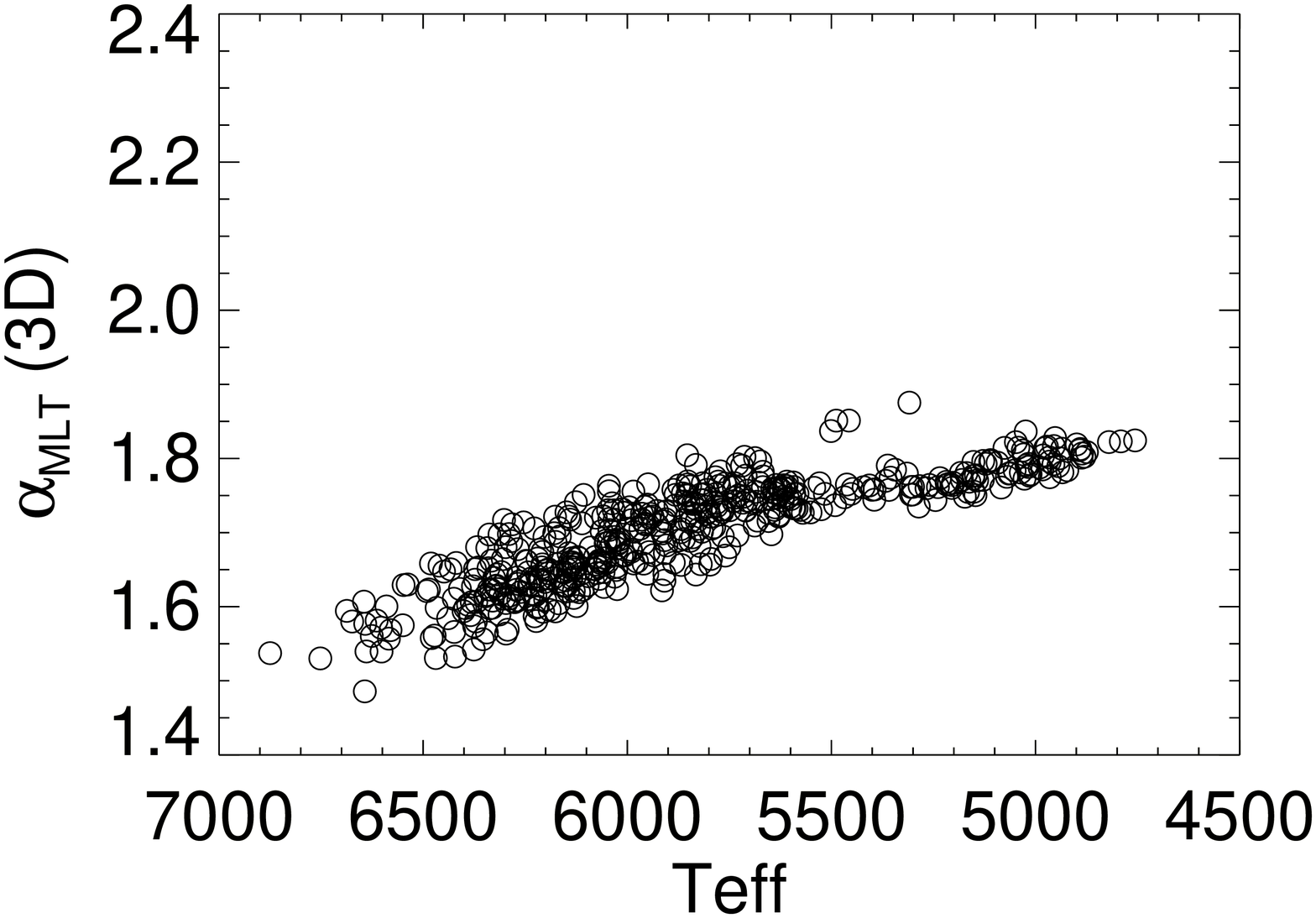}
\caption{Determination of $\almlt$ using different priors in \texttt{BeSPP}. Bottom right panel shows results directly obtained from $\almlt$ values based on 3D stellar atmospheres \cite{magic:2015}.
\label{fig:mlt}}
\end{figure}
%++++++++++++++

We use \texttt{BeSPP} to determine posterior values for $\almlt$ for our whole sample, using as before different assumptions for priors. Ideally, we would like to obtain good constraints for $\almlt$ in the case of flat priors that could be used to test theoretical determinations of $\almlt$ based on stellar atmosphere models. This has been attempted before, with different data and methods, by \cite{bonaca:2012}. Our initial results are shown in Fig.\,\ref{fig:mlt}, where $\almlt$ is shown as a function of $\teff$ and color-coded to represent $\yini$. For the Flat case, there is a clear correlation between $\almlt$ and $\yini$, regardless of $\teff$. 
Although it is not impossible that there is a physical dependence of $\almlt$ on the helium abundance of stars, i.e. on their composition, we are more inclined to believe that seismic data and the $\feh$ from \cite{chaplin:2014} are not enough to give too meaningful constraints on $\almlt$. The results change partially once the $\Delta_{YZ}$ prior is added, but still a correlation between $\almlt$ and $\yini$ seems to be present. Only after adding the \textbf{3DMLT} prior this correlation is almost completely removed (bottom left panel). The lower right panel shows the $\almlt$ values corresponding to our whole sample that correspond to \cite{magic:2015} 3D atmosphere models, directly determined from the stars $\teff$, $\feh$ and $\logg$ values using formulae given in that reference but with a shift applied so that $\almlt$ values match for the Sun. Clearly, the bottom two panels look quite similar, because our \textbf{3DMLT} prior is constructed from \cite{magic:2015} results. A more rigurous study is required, at this point, to better assess the possibilities that global seismology has to offer for determining $\almlt$ values and testing 3D hydrodynamic atmosphere models. Given the correlation with $\yini$, i.e. with the stellar composition, it might be crucial to use actual $\feh$ values in the analysis before firm conclusions can be drawn.

The last exercise we present here relates to determination $\yini$ from asteroseismic data, much as done above for $\almlt$, but present results in terms of the enrichement parameter $\Delta = \Delta Y / \Delta Z$, where $\Delta Y = \yini - Y_{\rm SBBN}$,  $Y_{\rm SBBN}=0.2485$ is the standard Big Bang Nucleosynthesis (SBBN) value and $\Delta Z= \zini$ because $Z_{\rm SBBN}=0$. Attempts to constrain $\Delta$ from asteroseismology, based on modeling stars based on individual frequencies or combinations, have been plagued with difficulties  linked to problematic determination of helium abundances. In fact, in many cases helium abundances are too low, below the SBBN value \cite{metcalfe:2014,silva:2015}, which makes us wary of the robustness of results for the whole sample. Asteroseismic modelling of stars using individual frequencies is known to introduce biases. This might also be the case when using frequency separation ratios. Is the situation different when global seismic quantities are used instead?

\begin{figure}[h]
\centering
\includegraphics[width=4.3cm,clip]{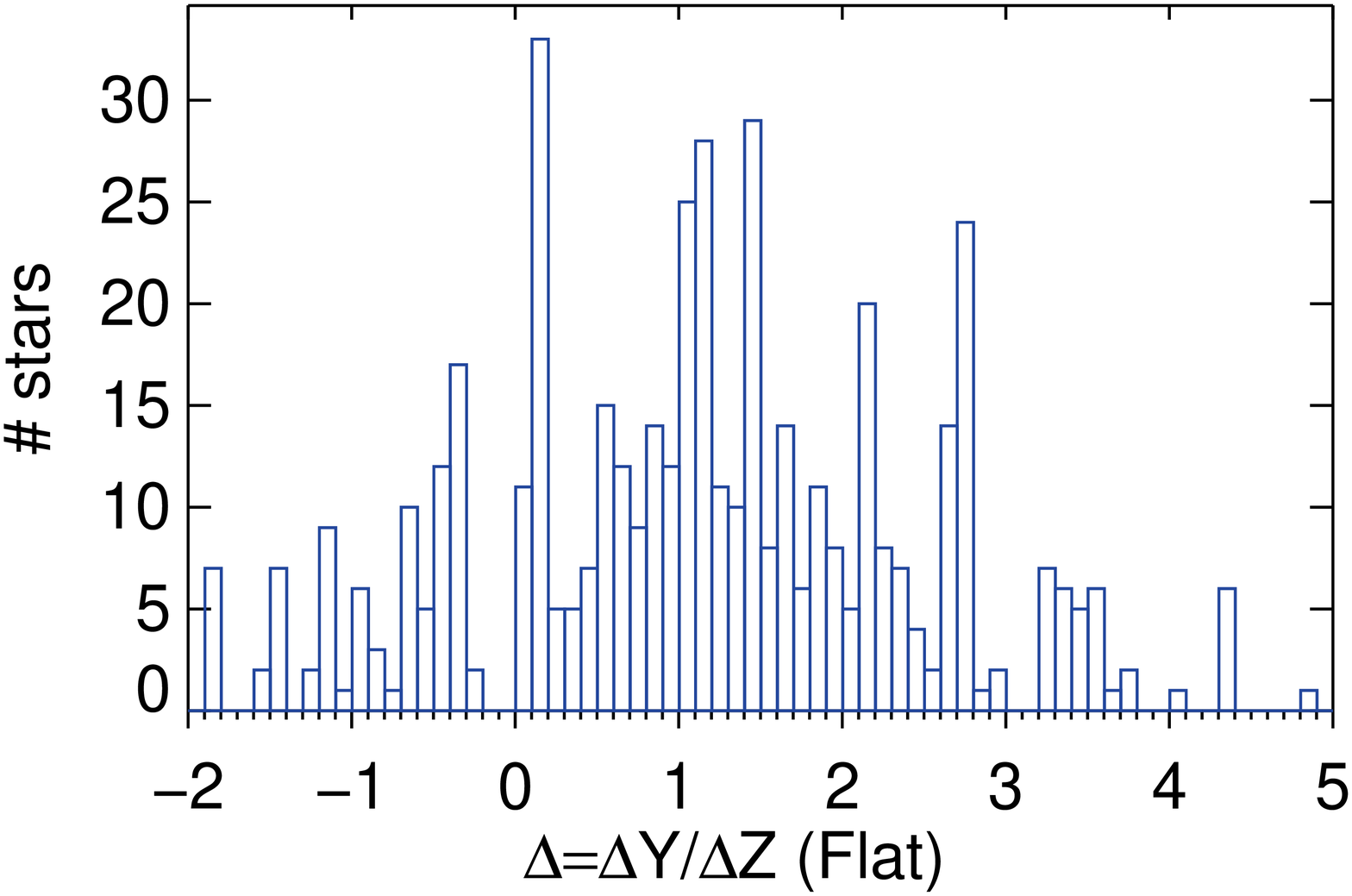}\includegraphics[width=4.3cm,clip]{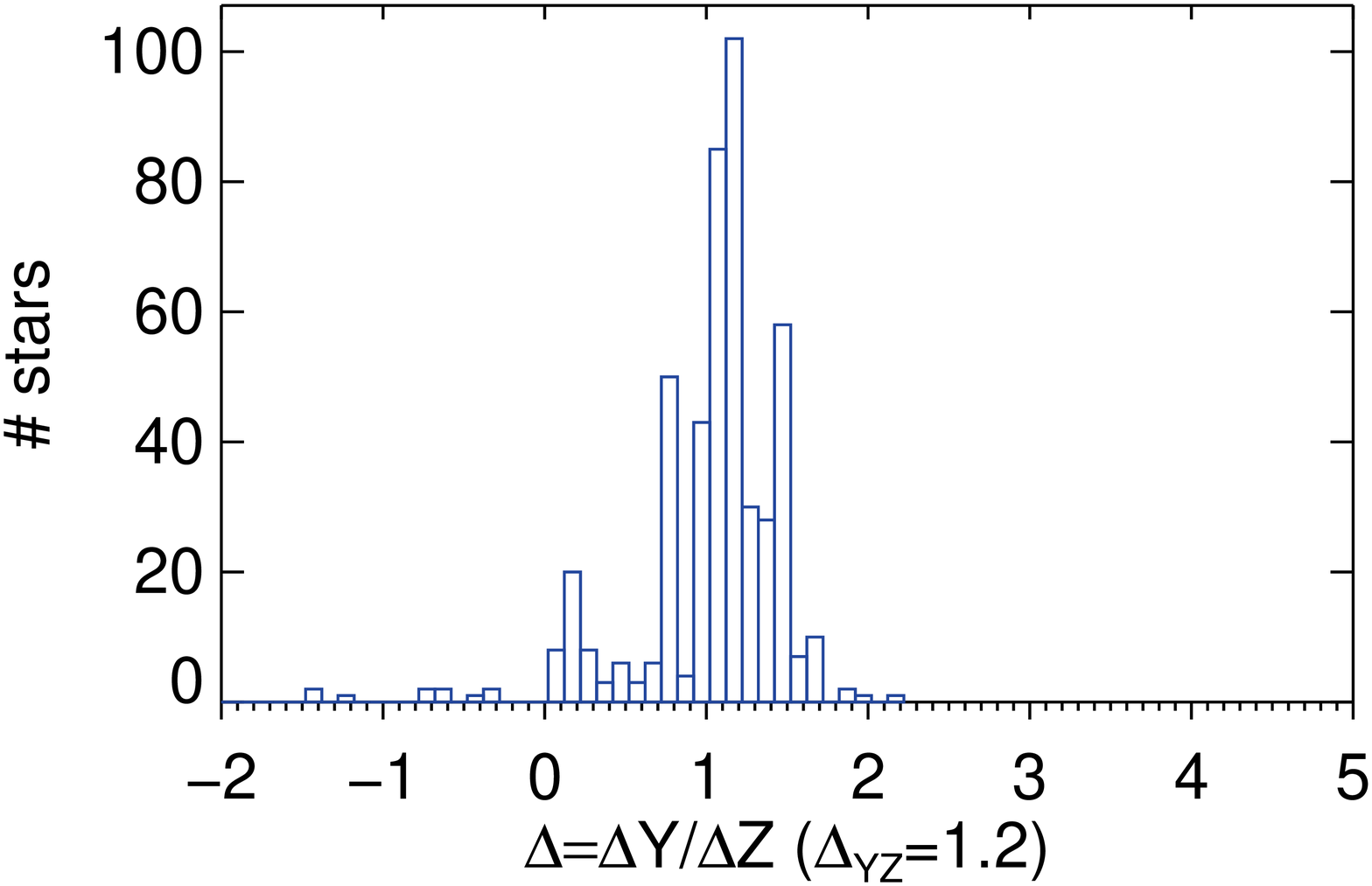}
\includegraphics[width=4.3cm,clip]{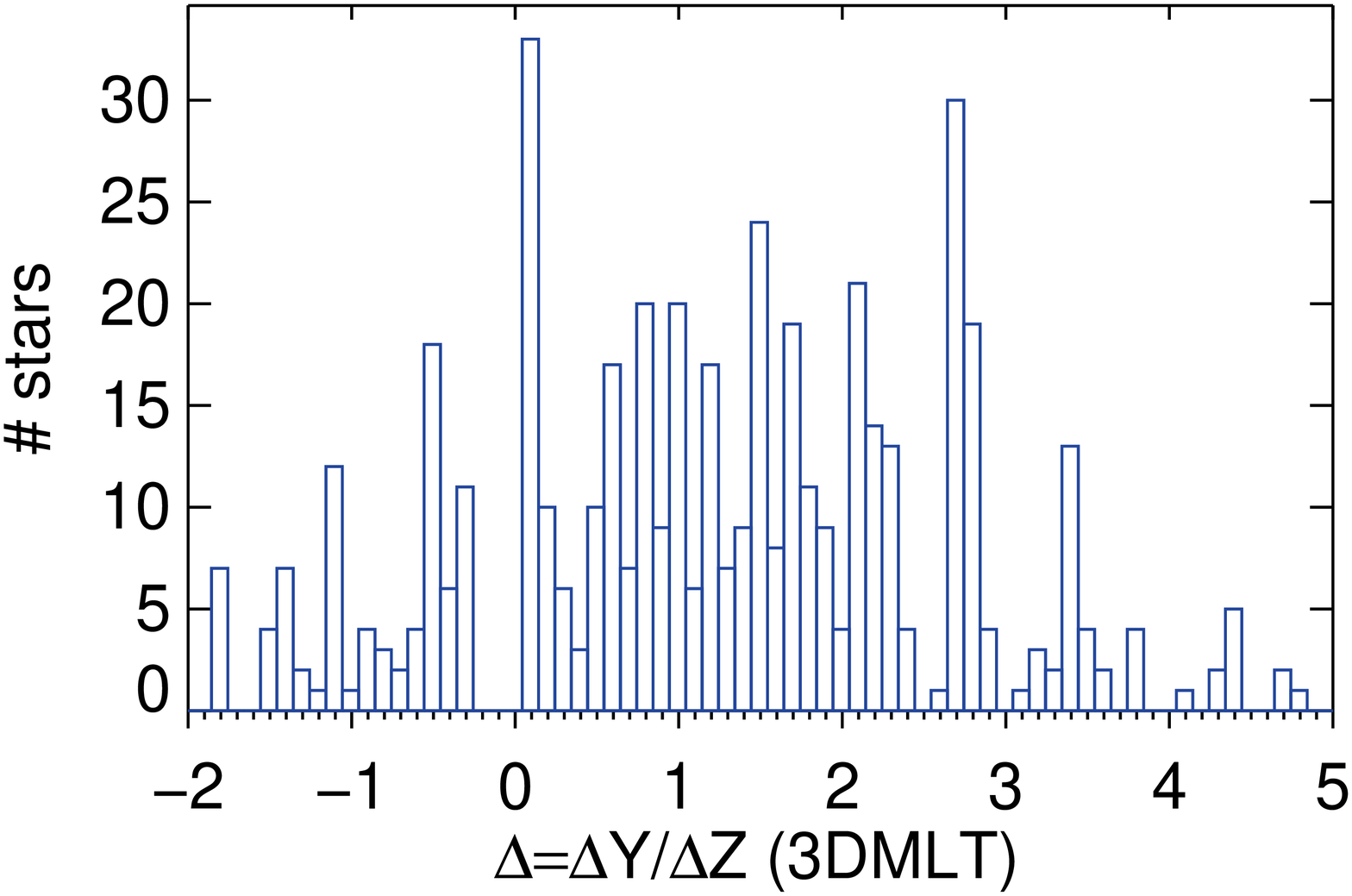}\includegraphics[width=4.3cm,clip]{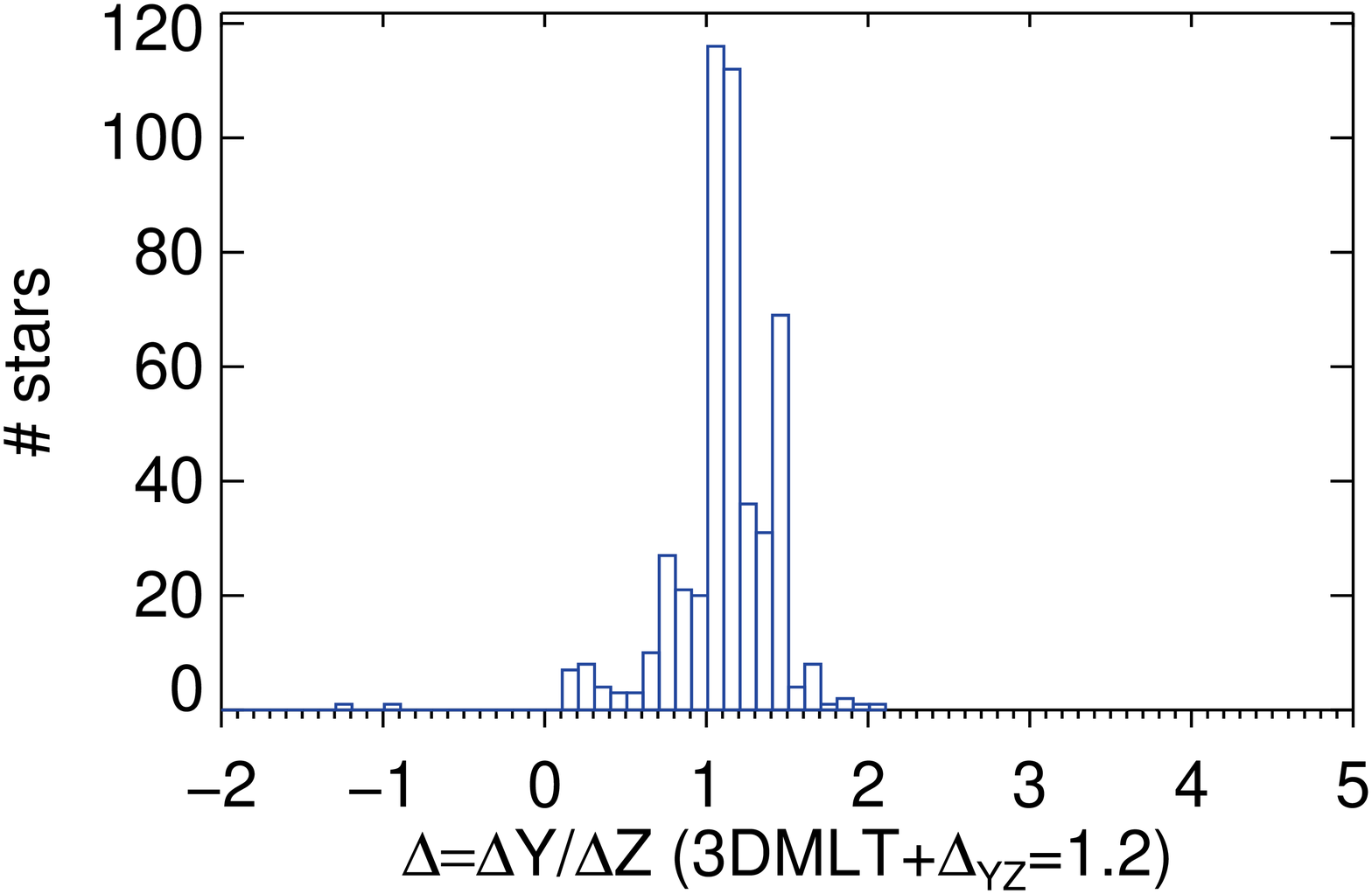}
\caption{Histograms showing the distribution of $\Delta = \Delta Y / \Delta Z$ for different assumptions regarding priors.}
\label{fig:dydz}       % Give a unique label
\end{figure}

Unfortunately, our first tests show than in fact $\yini$ values are also at odds with SBBN. Results for $\Delta$ are shown in Fig.\,\ref{fig:dydz} where, for different assumptions regarding priors, we show resulting $\Delta$ distributions. $\Delta < 0$ indicates $\yini < Y_{\rm SBBN}$. For the flat or the 3DMLT priors, clearly there are many stars that lead to low unrealistic $\yini$ values. Also, and despite the fact that most $\Delta$ values nucleate between 0.5,  the distribution looks very broad. When adding the $\Delta_{\rm YZ}$ prior, the $\Delta$ distribution is peaked around 1.2, but obviously this simply reflects the prior information. The combination of the $\Delta_{\rm YZ}$ and 3DMLT priors (lower right panel) gives the best behaved results, actually improving over the $\Delta_{\rm YZ}$-alone case (top right panel). Of course, $\Delta$ depends critically on $\zini$ and, as stated before, $\feh$ values in our sample are just fiducial values. Only after we improve this aspect, we can make more firm statements about this.

%++++++++++++++
%++++++++++++++

%-----------------------------------------------------------------------
\section{Conclusions}
\label{sec-con}

We have extended the capabilities of \texttt{BeSPP} to determine stellar physical parameters by adding $\yini$ and $\almlt$ as additional free parameters in its grid of stellar models. The fully new 5-dimensional grid of models relaxes the strong constraints imposed by: 1) the standard assumption that there is a one-to-one correlation between initial metal and helium abundance in stars; 2) that $\almlt$ takes a universal value that is solar calibrated. By adding priors in \texttt{BeSPP}, it is possible in principle to study different galactic chemical enrichment models, test $\almlt$ results from 3D model atmospheres and, very importantly, assess the impact of the usual assumptions (one-to-one $\yini-\zini$ relation and solar $\almlt$) in determination of stellar parameters such as mass, radius and age. 

In this preliminary exercise we have used \cite{chaplin:2014} sample, so we do not use reliable $\feh$ values. Therefore,results here are \emph{experimental}. However, one interesting and probably robust conclusion is that mass (and radius - not shown here) determinations using global seismic quantities is robust with respect to our ignorance on $\almlt$ and $\yini$.An estimate of the uncertainty in mass determinations yield a typical 4.4\% deviation between the two most extreme cases: flat priors on $\yini$ and $\almlt$ and priors that mimic typical GBM results (e.g. \cite{chaplin:2014}). This is an encouraging result. On the other hand, constraining $\yini-\zini$ relations or $\almlt$ values that can be used to test galactic chemical enrichment  or stellar atmosphere models seems at the moment a very daunting task because either $\yini$ or $\almlt$ can introduce changes in the model $\teff$ scale, leading to degenerate results. However, this needs further investigation, particularly by considering actual $\feh$ in the analysis.

%-----------------------------------------------------------------------
% BibTeX or Biber users please use (the style is already called in
% the class, ensure that the "woc.bst" style is in your local directory)
% \bibliography{name or your bibliography database}

\bibliographystyle{woc}

%***********************************************************************
\end{document}